\documentstyle[aps,epsf]{revtex}
\def\ie{ {i.e.,} }
\def\De{ {D} }

\def\d{{\rm d}}

\def\dt{{\delta \theta}}
\def\on{{\omega_n}}
\def\dt2{(\delta \theta)^2}

\def\p{\sigma}
\def\al{\alpha}
\def\be{\beta}

\def\de{\delta}

\def\im{{\rm i}}

\def\ch{\chi}

\def\lan{\left\langle}
\def\ran{\right\rangle}

\def\delx{\partial_x}

\def\nonum{\nonumber}

\def\dt{\frac{\partial}{\partial T}}

\def\e{{\rm e}}

\def\titheta{\tilde{\theta}}

\def\tiphi{\tilde{\phi}}

\def\virg{\;\;,}
\def\point{\;\;.}
\def\e{{\rm e }}
\def\nh{\hat{N}}
\def\tiD{\tilde{\Delta}}
\def\D{\Delta}
\def\tiE{\tilde{E}}
\def\jo #1#2#3#4{#1 {\bf #2}, #4  (#3)}  

\def\PRB{Phys.\ Rev.\ B}
\def\PRL{Phys.\ Rev.\ Lett.}

\def\SSC{Solid State Commun.}

\def\JPSJ{J.\ Phys.\ Soc.\ Jpn.}
\def\JSSC{J.\ Solid State Chem.}

\def\ADV{Adv.\ Phys.}

\def\PHC{Physica C}

\def\JLTP{J.\ Low Temp.\ Phys.}

\def\MCLC{Mol.\ Cryst.\ Liq.\ Cryst.}
\begin{document}
\draft
\title
{
Response Functions of Two-Coupled Chains of Tomonaga-Luttinger Liquids
}

\author
{
Hideo Yoshioka${}^{1,2}$
and Yoshikazu Suzumura${}^2$ \\
}

\address
{
${}^1$Department of Applied Physics, Delft University of Technology, 
Lorentzweg 1, 2628 CJ Delft, The Netherlands \\ 
${}^2$Department of Physics, Nagoya University, Nagoya 464-8602, Japan
} 

\date{Received \hspace*{3cm}}

\maketitle

\begin{abstract}
Properties of fluctuations 
 in two chains of Tomonaga-Luttinger liquids 
coupled by the interchain hopping 
 have been studied by calculating  
 retarded  response functions 
 $\chi^R_{\rho} (q_x,q_y;\omega)$ for charge and   
 $\chi^R_{\p} (q_x,q_y;\omega)$ for spin 
where 
 $q_x$ and $q_y (=0$  or $\pi$) denote 
   the longitudinal and transverse wave vector, respectively,   
 and $\omega$ is the frequency. 
 We have found the notable fact that 
 the repulsive intrachain interaction 
  results in the clear enhancement of  
    ${\rm{Im}}\chi^R_\p (q_x,\pi;\omega)$ and  
  the suppression of  ${\rm{Im}}\chi^R_\rho (q_x,\pi;\omega)$ 
   at  low energies.   
 This result indicates the importance  of the dynamical effect 
 by   the spin fluctuation  with $q_y = \pi$ and small $\omega$, 
 which has a possibility to give rise to the  attractive interaction
for the electron pairing. 
\end{abstract}

\vspace{2em}

\hspace{4.5em}PACS numbers: 71.10.Hf, 71.10.Pm, 74.20.Mn
\pacs
{
{{\it keywords}} : Tomonaga-Luttinger liquids, two chains, 
response functions, superconductivity
}
\section{Introduction}

Two-chain system of Tomonaga-Luttinger liquids 
coupled by the interchain hopping and/or the exchange interaction 
 is a basic model which connects
  one-dimensional interacting electron systems 
with quasi-one- and two-dimensional systems\cite{Dagotto-Rice-Rev}. 
 When the intrachain interaction  between conduction electrons 
is repulsive,  
   the ground state  
 exhibits  the superconducting one
\cite{Dagotto-Riera-Scalapino,Fabrizio-Parola-Tosatti,Fabrizio,Shimoi-Yamaji-Yanagisawa,Nagaosa,Schulz,Balents-Fisher,Shelton-Tsvelik}
 where  the spin gap found in the half-filling case 
\cite{Dagotto-Riera-Scalapino,Johnston-Johnston,Hiroi,Senachal} 
still survives. 
 The result is of interest for understanding 
  the experimental fact that 
 superconducting states have been observed in  
 the doped ladder systems, 
 ${\rm Sr_{0.4} Ca_{13.6} Cu_{24} O_{41.48} }$, 
 under the pressure\cite{Uehara-Nagata-Akimitsu-Takahashi-Mori-Kinoshita}.  
Further, the phase transition 
of quasi-one-dimensional organic conductors, 
${\rm (TMTSF)_2 X}$, which takes place 
from spin density wave state to superconducting one 
with increasing pressure\cite{Jerome,TMTSF-rev} 
could be related with crossover in the two chains\cite{Yoshioka-Suzumura-4}.

It has been known that the superconductivity 
 in the two-chain system 
results from the repulsive intrachain interaction 
of the backward scattering between electrons with anti-parallel spin
\cite{Schulz,Yoshioka-Suzumura-4} 
when the interchain hopping is relevant\cite{relevant-condition}. 
 The study of the spin and charge fluctuations    
 in terms of the response functions 
 clarifies the dynamics of the two-coupled chains, 
and is useful  
for understanding 
 the origin of  superconductivity 
 not only for   two-chain systems but also for  
 quasi-one-dimensional conductors.  
 The response functions have been examined for 
 two-coupled chain systems  with  only the forward 
 scattering
\cite{Yoshioka-Suzumura-5}, 
 but  those  in the presence of the backward scattering 
are not yet clear.   
 In the present  paper, 
 by use of effective Hamiltonian which is based on  
 the renormalization group analysis,
 we investigate  retarded  response functions for charge and spin, 
 given by  $\chi^R_{\rho} (q_x,q_y;\omega)$ and   
 $\chi^R_{\p} (q_x,q_y;\omega)$ where   
 $q_x$ and $q_y (=0$  or $\pi$)  are
   the longitudinal and transverse wave vector, respectively,   
 and $\omega$ is the frequency. 
 We demonstrate  that 
 the spin fluctuation dominates 
 the low-lying excitation with $q_y = \pi$
 from the calculation of the $\omega$-dependence of both 
 ${\rm Im} \chi^R_\rho(q_x,\pi;\omega)$ and 
 ${\rm Im} \chi^R_\p(q_x,\pi;\omega)$.  

In section II,  the  Hamiltonian for the two-coupled chain is 
 represented by the phase variable 
 and is  investigated in terms of the new Fermion fields. 
The results of the calculation of the retarded response functions, 
 $\chi^R_{\rho} (q_x,q_y;\omega)$ and   
 $\chi^R_{\p} (q_x,q_y;\omega)$, are shown in section III. 
Section IV is devoted to summary and discussion. 
  
\section{Model}
 We consider a model of  two chains  coupled by  the interchain hopping $t$ where 
  each chain consists of conduction electrons 
 with repulsive intrachain 
   interactions of both the backward scattering 
 and the forward scattering.   
 We note that 
  the single chain in the absence of    $t$    
leads to Tomonaga-Luttinger liquids
\cite{Solyom}. 
 The present system can be expressed  by the phase Hamiltonian
\cite{Schulz,Yoshioka-Suzumura-4},        
  which is based on the bosonization method. 
 In case  that   the energy is 
  less than $t$,
 the present  Hamiltonian 
 is obtained by neglecting 
 the non-linear terms including the misfit parameter 
 $2q_0 = 4t/v_F$ which is originated from 
the separation of Fermi wave vector due to the 
interchain hopping\cite{Schulz,Yoshioka-Suzumura-4,notation}.       
 Thus our Hamiltonian is given as,        
\begin{eqnarray}
	{\cal H} & = & \frac{v_\theta}{4\pi} \int \d x 
	\left\{ \frac{1}{\eta_\theta} (\delx \theta_+)^2 
	      + \eta_\theta (\delx \theta_-)^2 \right\} 
	 +  \frac{v_\phi}{4\pi} \int \d x 
	\left\{ \frac{1}{\eta_\phi} (\delx \phi_+)^2 
	      + \eta_\phi (\delx \phi_-)^2 \right\} \nonum \\
	& + & \frac{v_F}{4\pi} \int \d x 
	\left\{ (\delx \titheta_+)^2 + (\delx \titheta_-)^2 \right\} 
	 +  \frac{v_F}{4\pi} \int \d x	       
        \left\{ (\delx \tiphi_+)^2 + (\delx \tiphi_-)^2 \right\} \nonum \\ 
	& + & \frac{v_F}{\pi \al'^2} 
	      \int \d x \biggl[
	      g_-  \cos \sqrt{2} \titheta_-  \cos \sqrt{2} \tiphi_- 
     +  
        g_+ \cos \sqrt{2} \titheta_-  \cos \sqrt{2} \tiphi_+ \nonum 
\\ 	
	&  & + 
      g^*_1 \cos \sqrt{2} \phi_+ 
     \left(   \cos \sqrt{2} \tiphi_+ 
            + \cos \sqrt{2} \tiphi_- 
            - \cos \sqrt{2} \titheta_-  
             \right)   \biggr]  \virg   
	\label{eqn:hl}  
\end{eqnarray}  
 where $\al' \sim v_F / t $ and $v_F$ is the Fermi velocity. 
 The phase variables 
 $\theta_\pm$ ($\phi_\pm$) and $\titheta_\pm$ ($\tiphi_\pm$) express
 the fluctuations of the total charge (spin) and the transverse charge 
 (spin) where 
$ 
[ \theta_+ (x), \theta_-(x')] = 
[ \titheta_+ (x), \titheta_-(x')] = 
[ \phi_+ (x), \phi_-(x')] = 
[ \tiphi_+ (x), \tiphi_-(x')] = 
 i \pi {\rm sgn}(x-x')
$. 
 The parameter $g_1$ ($g_2$) is a matrix element of the 
 backward (forward) scattering, whose conventional definition is 
 given by $g_i \to g_i/(2\pi v_F)$\cite{Solyom}. 
 The coefficients 
 in Eq.(\ref{eqn:hl}) are given as  
 $v_\theta = v_F \sqrt{( 1 + 2 g_2 - g_1 )( 1 - 2 g_2 + g_1 )}$,  
 $v_\phi = v_F \sqrt{( 1 - g^*_1 )( 1 + g^*_1 )}$, 
 $\eta_\theta = \sqrt{( 1 - 2 g_2 + g_1 )/( 1 + 2 g_2 - g_1 )}$, 
 $\eta_\phi = \sqrt{( 1 + g^*_1 )/( 1 - g^*_1 )}$ and  
 $g_\pm = g_2 - g_1/2 \pm g^*_{1}/2$ 
\cite{Schulz}       
 where  $g_1^* = g_1/\big\{ 1 + 2 g_1 \ln (v_F/t\al) 
 \big\}$ \cite{Solyom} 
and  $v_F \al^{-1}$ is of the order of Fermi energy. 

 In Eq.(\ref{eqn:hl}), 
 the excitation spectrum of 
 the total charge  fluctuation is gapless  and 
 has  the same velocity as 
  that in the absence of the interchain hopping.
 On the other hand, the interchain hopping plays important roles 
in the other degree of freedoms 
since the Hamiltonian expressing these fluctuation 
 includes  complex non-linear terms.
   For  $g_2 - g_1/2 > 0$,   
 the renormalization to the strong coupling has been obtained  
 for the terms proportional to 
 $\cos \sqrt{2} \titheta_-  \cos \sqrt{2} \tiphi_+$,  
 $\cos \sqrt{2} \phi_+ \cos \sqrt{2} \tiphi_+$ and  
 $\cos \sqrt{2} \phi_+ \cos \sqrt{2} \titheta_-$\cite{Yoshioka-Suzumura-4}.  
 Thus the quantities, $\cos \sqrt{2} \titheta_-$, 
 $\cos \sqrt{2} \tiphi_+$ and $\cos \sqrt{2} \phi_+$ have 
finite expectation values, and 
 the gapful excitations appear in the total spin,   
 the transverse charge and the transverse spin degrees of freedoms. 
 In case of the strong coupling, 
 the behavior of fluctuations 
with low energy  can be well described by 
use of the effective Hamiltonian in which  the non-linear term 
with finite expectation value can be treated  
 as  the  averaged one\cite{Schulz,Finkelstein-Larkin}.   
 Following such a procedure,  
 the parts describing the gapful excitation in Eq.(\ref{eqn:hl}) 
is rewritten as
\begin{eqnarray}
	{\cal H'} & = & 
	\frac{v_\phi}{4\pi} \int \d x 
	\left\{ \frac{1}{\eta_\phi} (\delx \phi_+)^2 
	      + \eta_\phi (\delx \phi_-)^2 \right\} \nonum \\
	& + & \frac{v_F}{4\pi} \int \d x 
	\left\{ (\delx \titheta_+)^2 + (\delx \titheta_-)^2 \right\} 
	 +  \frac{v_F}{4\pi} \int \d x	       
        \left\{ (\delx \tiphi_+)^2 + (\delx \tiphi_-)^2 \right\} \nonum \\ 
	& + & \frac{v_F}{\pi \al'^2} g_+ 
	      \int \d x \biggl[
        \lan \cos \sqrt{2} \titheta_- \ran  \cos \sqrt{2} \tiphi_+ 
        + \cos \sqrt{2} \titheta_-  \lan  \cos \sqrt{2} \tiphi_+ \ran \biggr] \nonum \\
	& + & \frac{v_F}{\pi \al'^2} g^*_1 
	      \int \d x \biggl[ 
        \lan \cos \sqrt{2} \phi_+ \ran \cos \sqrt{2} \tiphi_+ 
        + \cos \sqrt{2} \phi_+  \lan \cos \sqrt{2} \tiphi_+ \ran \nonum \\ 
        & & \hspace{13ex}
            - \lan \cos \sqrt{2} \phi_+ \ran  \cos \sqrt{2} \titheta_-  
            - \cos \sqrt{2} \phi_+ \lan  \cos \sqrt{2} \titheta_- \ran 
                \biggr]  \virg   
	\label{eqn:hl2}  
\end{eqnarray}  
where $\lan \cdots \ran$ expresses the  thermal average.
  In Eq.(\ref{eqn:hl2}), the terms proportional to    
$\lan \cos \sqrt{2} \titheta_- \ran  \cos \sqrt{2} \tiphi_-$ and 
$\lan \cos \sqrt{2} \phi_+ \ran  \cos \sqrt{2} \tiphi_-$
 have been discarded 
 because these terms gives rise to the finite expectation value of 
 $\cos \sqrt{2} \tiphi_-$, which is inconsistent  
  with the results derived from the renormalization group analysis. 
 The quantities, $\lan \cos \sqrt{2} \titheta_- \ran$, 
 $\lan \cos \sqrt{2} \tiphi_+ \ran$ and 
$\lan \cos \sqrt{2} \phi_+ \ran$ are related to the gap 
 $\tiD_s$, $\tiD_c$ and $\Delta_s$ as 
\begin{eqnarray}
\tiD_s &=& \frac{v_F}{\al'}
\left\{ g_+ \lan \cos \sqrt{2} \titheta_- \ran  
      + g_1^* \lan \cos \sqrt{2} \phi_+ \ran \right\} \virg 
\label{eqn:gap1}\\
\tiD_c &=& \frac{v_F}{\al'}
\left\{ g_+ \lan \cos \sqrt{2} \tiphi_+ \ran  
      - g_1^* \lan \cos \sqrt{2} \phi_+ \ran \right\} \virg
\label{eqn:gap2}\\
\Delta_s &=& \frac{v_F}{\al'} g_1^*
\left\{\lan \cos \sqrt{2} \tiphi_+ \ran  
      - \lan \cos \sqrt{2} \titheta_- \ran \right\} \virg 
\label{eqn:gap3} 
\end{eqnarray} 
and are determined self-consistently 
 similar to the previous case  
  $g_1 = 0$\cite{Yoshioka-Suzumura-5}.

 Now we examine  Eq.(\ref{eqn:hl2}) 
 by introducing   the  Fermionic representation  
 \cite{Luther-Emery} which has been applied 
 for $\eta_\phi =1$. 
 Although  the case $\eta_\phi =1$  is obtained 
 as a special case in the  calculation of 
the renormalization equation  \cite{Yoshioka-Suzumura-4}, 
  it is considered that 
 such a setting does not change  qualitatively 
the behavior of the solutions.   
 The effect of $\eta_\phi \neq 1$ is discussed 
 in the last section. 
 Thus Eq.(\ref{eqn:hl2}) is expressed as
\begin{eqnarray}
{\cal H}' &=& v_F \int \d x 
\left\{ 
\psi^\dagger_1 (-\im \delx \psi_1) - \psi^\dagger_2 (-\im \delx \psi_2)
\right\}
+ \tiD_s \int \d x 
\left\{
\psi^\dagger_2 \psi_1 + \psi^\dagger_1 \psi_2
\right\}
\nonum \\
&+& v_F \int \d x 
\left\{ 
\psi^\dagger_3 (-\im \delx \psi_3) - \psi^\dagger_4 (-\im \delx \psi_4)
\right\}
+ \tiD_c \int \d x 
\left\{
\im \psi_3 \psi_4 - \im \psi^\dagger_4 \psi^\dagger_3
\right\}
\nonum \\
&+& v_\phi \int \d x 
\left\{ 
\psi^\dagger_5 (-\im \delx \psi_5) - \psi^\dagger_6 (-\im \delx \psi_6)
\right\}
+ \Delta_s \int \d x 
\left\{
\psi^\dagger_6 \psi_5 + \psi^\dagger_5 \psi_6
\right\} \virg
\label{eqn:hdmf}
\end{eqnarray}
 where  
$\psi_1$ ($\psi_2$), $\psi_3$ ($\psi_4$) and $\psi_5$ ($\psi_6$)
are the field operators of right going (left going) Fermions 
 corresponding to the transverse spin, the transverse charge 
and the total spin degree of freedoms, respectively. 
 In terms of  $\tiphi_\pm$, $\titheta_\pm$ and $\phi_\pm$, 
  field operators, 
 $\psi_j$ 
 are  defined by 
\begin{eqnarray}
\psi_{1+n} &=& \frac{1}{\sqrt{2\pi \al'}}
           \e^{\frac{\im}{\sqrt{2}}((-1)^n \tiphi_+ + \tiphi_-)}
           \e^{\im (-1)^n \frac{\pi}{2} (\nh_1 + \nh_2)} \nonum \virg
\\ 
\psi_{3+n} &=& \frac{1}{\sqrt{2\pi \al'}}
           \e^{\frac{\im}{\sqrt{2}}((-1)^n \titheta_+ + \titheta_-)}
           \e^{\im (-1)^n  \frac{\pi}{2} (\nh_3 + \nh_4) + \im \pi (\nh_1 + \nh_2)} \nonum \virg
\\
\psi_{5+n} &=& \frac{1}{\sqrt{2\pi \al'}}
           \e^{\frac{\im}{\sqrt{2}}((-1)^n \phi_+ + \phi_-)}
           \e^{\im (-1)^n  \frac{\pi}{2} (\nh_5 + \nh_6) 
           + \im \pi (\nh_1 + \nh_2 + \nh_3 + \nh_4)} 
            \virg
\label{eqn:field}
\end{eqnarray}
where 
$n =$ 0 and 1, and   
$\nh_i$ ($i = 1 \sim 6$) is a number operator 
 of the $i$-th Fermion.  
The Hilbert space is  taken so that  
 the numbers,  $N_1 + N_2$, $N_3 + N_4$ and $N_5 + N_6$, 
 are  even integers
( $N_i$ ($i = 1 \sim 6$) : eigenvalue of $\nh_i$ ) 
in deriving Eq.(\ref{eqn:hdmf}). 
Note that phase variables, 
$\tiphi_\pm$, $\titheta_\pm$ and $\phi_\pm$ can be also 
 expressed as, 
\begin{eqnarray}
\tiphi_{\pm}(x) &=&  
-  \sum_{q \not= 0 } \frac{\sqrt{2} \pi \im}{q L} 
\e^{(- \al' |q|/2 + \im q x)} 
 \left\{ \De_1(-q) \pm  \De_2(-q) \right\} \virg
\label{eqn:tiphi}\\
\titheta_{\pm}(x) &=&  
 -  \sum_{q \not= 0 } \frac{\sqrt{2} \pi \im}{q L} 
\e^{(- \al' |q|/2 + \im q x)} 
 \left\{ \De_3(-q) \pm  \De_4(-q) \right\} \virg 
\label{eqn:titheta}\\
\phi_{\pm}(x) &=&  
 -  \sum_{q \not= 0 } \frac{\sqrt{2} \pi \im}{q L} 
\e^{(- \al' |q|/2 + \im q x)} 
 \left\{ \De_5(-q) \pm  \De_6(-q) \right\} \virg
\label{eqn:phi}
\end{eqnarray}
where
$ 
 \De_j(-q) = \int \psi_j^{\dagger} \psi_j \e^{- {\rm i} q x} \d x \point
$

\begin{figure}
\centerline{\epsfysize=10.0cm\epsfbox{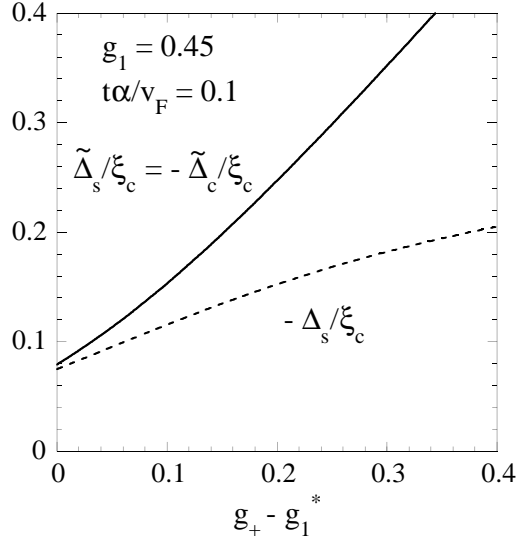}}
\vspace{-2.2cm}
\caption{
 Solutions of the gap equations of 
 Eqs.(11)$-$(13) as a function of $g_+ - g_1^*$,   
for $g_1 = 0.45$ and $t \alpha / v_F = 0.1$.  
The solid line and the dotted one express 
$\tiD_s/\xi_c = - \tiD_c/\xi_c$ and $- \Delta_s/\xi_c$, respectively,
where $\xi_c$ is the cut-off energy of the order of $t$.   
}  
\label{fig:1}
\end{figure}

 In Eq.(\ref{eqn:hdmf}), the excitation spectra of the  total spin, 
the transverse charge 
and the transverse spin degrees of freedom are respectively 
 calculated as 
$E_{k,s} = \sqrt{\xi_{k,s}^2 + \D_s^2}$, 
$\tiE_{k,c} = \sqrt{\xi_{k}^2 + \tiD_c^2}$ and 
$\tiE_{k,s} = \sqrt{\xi_{k}^2 + \tiD_s^2}$, where
$\xi_{k,s} = v_\phi k$ and $\xi_k = v_F k$.  
 By use of Eq.(\ref{eqn:hdmf}),  
the self-consistent equations 
Eqs.(\ref{eqn:gap1}) - (\ref{eqn:gap3}) 
for 
 $\tiD_s$, $\tiD_c$ and $\Delta_s$  
 are rewritten as 
\begin{eqnarray}
\frac{\tiD_s}{\pi v_F} &=& 
g_+ 
\left( 
\im \langle \psi_3 \psi_4 \rangle - \im \langle \psi_4^\dagger \psi_3^\dagger \rangle 
\right)
+ g_1^* 
\left( 
\langle \psi_6^\dagger \psi_5 \rangle + \langle \psi_5^\dagger \psi_6 \rangle 
\right)
\nonum \\
&=& -g_+ \frac{\tiD_c}{L} \sum_k \frac{1}{\tiE_{k,c}}
- g_1^* \frac{\D_s}{L} \sum_k \frac{1}{E_{k,s}} \virg
\label{eqn:tids}
\\
\frac{\tiD_c}{\pi v_F} &=& 
g_+ 
\left( 
\langle \psi_2^\dagger  \psi_1 \rangle + \langle \psi_1^\dagger  \psi_2 \rangle 
\right)
- g_1^* 
\left( 
\langle \psi_6^\dagger \psi_5 \rangle + \langle \psi_5^\dagger \psi_6 \rangle 
\right)
\nonum \\
&=& - g_+ \frac{\tiD_s}{L} \sum_k \frac{1}{\tiE_{k,s}}
+ g_1^* \frac{\D_s}{L} \sum_k \frac{1}{E_{k,s}} \virg
\label{eqn:tidc} 
\\
\frac{\Delta_s}{\pi v_F} &=& g_1^* 
\left( 
\langle \psi_2^\dagger \psi_1 \rangle 
+  \langle \psi_1^\dagger \psi_2 \rangle   
-  \im \langle \psi_3 \psi_4 \rangle 
+  \im \langle \psi_4^\dagger \psi_3^\dagger \rangle 
\right) 
\nonum \\
&=& g_1^* 
\left\{
- \frac{\tiD_s}{L} \sum_k \frac{1}{\tiE_{k,s}}
+ \frac{\tiD_c}{L} \sum_k \frac{1}{\tiE_{k,c}} 
\right\} \point
\label{eqn:ds}
\end{eqnarray}
 By noting that the  gap equations lead to  $\tiD_s = - \tiD_c 
 \equiv \tiD $ 
 due to  $g_+ > 0$,  
  we use $\tiE_k = \sqrt{\xi_k^2 + \tiD^2} = \tiE_{k,s} = \tiE_{k,c}$ 
in the following.  
 In Fig.\ref{fig:1},  the numerical results of the gap equations 
 are shown  
 with a choice of  $g_1 = 0.45$ and   $t \al/v_F = 0.1$ 
 where   $g_+ > g_1^*$. 
The result of $|\tiD_s| = |\tiD_c| > |\Delta_s|$ 
obtained in Fig.\ref{fig:1} 
is consistent with that of 
the renormalization analysis\cite{Yoshioka-Suzumura-4} 
 showing the fact that  the term proportional to 
$\cos \sqrt{2} \titheta_-  \cos \sqrt{2} \tiphi_+$ 
is scaled to the  strong coupling regime  
 faster than the other relevant terms. 

\section{Response Functions}
We calculate 
the response functions defined by  ($\nu$ = $\rho$ and $\sigma$ )
\begin{eqnarray}
\ch^R_\nu (q_x, q_y; \omega) 
&=&  \frac{1}{2}
\int^\be_0 \d \tau \int \d (x - x') \e^{\im \on \tau} \e^{-\im q_x (x - x') }
\nonum \\ 
 & &  \times 
         \lan T_\tau 
	 \left\{ \nu(x, 1 ; \tau) + \e^{\im q_y} \nu(x, 2 ; \tau) \right\}
	 \left\{ \nu(x', 1 ; 0) + \e^{\im q_y} \nu(x', 2 ; 0) \right\}
	 \ran \Bigg{|}_{\im \on \to \omega + \im \de} \virg
\label{eqn:kainu} 
\end{eqnarray}
where 
$\rho(x,i;\tau)$ and $\p(x,i;\tau)$ denote
 operators of the charge and spin densities  
at the  $i$(=1,2)-th chain, respectively. 
 In Eq.(\ref{eqn:kainu}), 
  the longitudinal wave vector $q_x$  is much smaller than $2k_F$
 ( $k_F$ is the Fermi wave vector ) 
 and  $q_y (= 0, \pi)$ is  the transverse wave vector. 
Therefore the operators, 
$\nu(1) \pm \nu(2)$ 
($\equiv \nu(x,1;\tau) \pm \nu(x,2;\tau)$), 
in Eq.(14) 
are given as 
$\rho(1) + \rho(2) = \partial_x \theta_+/\pi$, 
$\p(1) + \p(2) = \partial_x \phi_+/\pi$, 
$\rho(1) - \rho(2) = \{1/(2\pi \al')\} \sum_{p,\p,\mu}
p \mu {\rm e}^{\im p \mu q_0 x} 
\exp\{-\im p \mu (\titheta_+ + p \titheta_-)/\sqrt{2}\}
\exp\{-\im p \mu \p (\tiphi_+ + p \tiphi_-)/\sqrt{2}\}$ 
and 
$\p(1) - \p(2) = \{1/(2\pi \al')\} \sum_{p,\p,\mu}
p \mu \p {\rm e}^{\im p \mu q_0 x} 
\exp\{-\im p \mu (\titheta_+ + p \titheta_-)/\sqrt{2}\}
\exp\{-\im p \mu \p (\tiphi_+ + p \tiphi_-)/\sqrt{2}\}$
where the summation denotes $p=\pm$, $\p=\pm$ and $\mu = \pm$
\cite{Yoshioka-Suzumura-5}. 
The charge response functions   
$\chi^R_\rho(q_x, 0; \omega)$ with $q_y = 0$, which 
 is  evaluated  straightforwardly from the 
 total charge density given by 
$\partial_x \theta_+/\pi$,  
 is the same as that in the absence of the interchain hopping.  
On the other hand, 
$\chi^R_\p (q_x, 0; \omega)$, $\chi^R_\rho (q_x, \pi; \omega)$ and 
$\chi^R_\p (q_x, \pi; \omega)$  are calculated  from 
  the Fermionic representation 
 of Eq.(\ref{eqn:field}). 
 These response functions at absolute zero temperature are given as
\begin{eqnarray}
\chi^R_\rho (q_x, 0 ; \omega) 
&=&
\frac{\eta_\theta q_x}{\pi} 
\left\{
  \frac{1}{v_\theta q_x + \omega + \im \de} 
+ \frac{1}{v_\theta q_x - \omega - \im \de} 
\right\} \virg 
\label{eqn:chiro0} \\  
\chi^R_\p (q_x, 0 ; \omega) 
&=& 
\frac{1}{L} \sum_k 
\left(
\frac{1}{E_{k,s} + E_{k+q_x,s} + \omega + \im \de} 
+ \frac{1}{E_{k,s} + E_{k+q_x,s} - \omega - \im \de}   
\right) \nonum \\
& & \times
\left(
1 - \frac{\xi_{k,s} \xi_{k+q_x,s}}{E_{k,s}E_{k+q_x,s}} 
- \frac{\D_s^2}{E_{k,s}E_{k+q_x,s}}
\right) \virg
\label{eqn:chi0}
\\
\chi^R_{\rho(\p)} (q_x,\pi;\omega) 
&=&
\frac{1}{2L} \sum_k \sum_{\nu = \pm} 
\left(
\frac{1}{\tiE_k + \tiE_{k'} - \omega - \im \de}
+ \frac{1}{\tiE_k + \tiE_{k'} + \omega + \im \de}
\right) \nonum \\
& & \times
\left(
1 + \frac{\xi_k}{\tiE_k} \frac{\xi_{k'}}{\tiE_{k'}} 
+ (-)  \frac{\tiD_c}{\tiE_k} \frac{\tiD_s}{\tiE_{k'}}
\right)_{{k'} = \nu q_x + q_0 - k} \point
\label{eqn:chipi}
\end{eqnarray}
Note that Eqs. (\ref{eqn:chi0}) and (\ref{eqn:chipi}) have been derived  
in terms of Eq.(\ref{eqn:phi}) and Eqs.(\ref{eqn:field}), respectively.    

\begin{figure}
\centerline{\epsfysize=12.5cm\epsfbox{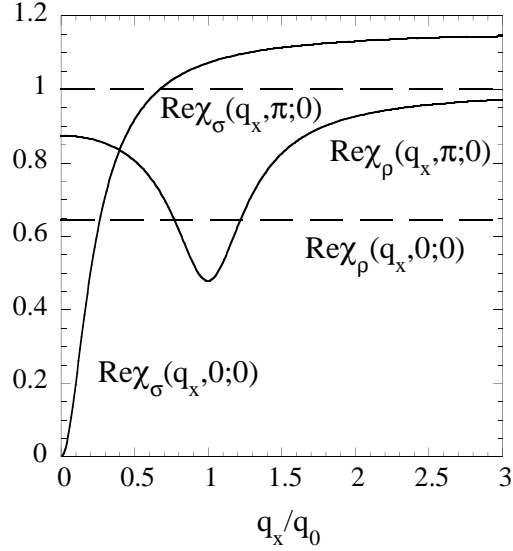}}
\vspace{-4.7cm}
\caption{
 The $q_x/q_0$-dependence of  
 the real part of the response functions, 
 ${\rm Re} \chi_\rho(q_x,0;0)$, 
 ${\rm Re} \chi_\p(q_x,0;0)$, 
 ${\rm Re} \chi_\rho(q_x,\pi;0)$ and  
 ${\rm Re} \chi_\p(q_x,\pi;0)$
 which are normalized by $2/\pi v_F$.  
 Here the parameters are chosen as $|\tiD|/2t = 0.124$ 
and $|\tiD_s|/2t = 0.076$ which corresponds to
 $g_1 = 0.45$, $g_2 = 0.5$ and $t \alpha / v_F = 0.1$.
}
\label{fig:2}
\end{figure}
For $\omega$ = 0, 
 the real parts  of Eqs.(\ref{eqn:chiro0})-(\ref{eqn:chipi}) 
 can be calculated analytically.  
 The quantities,  ${\rm{Re}} \chi_\rho (q_x,0;0)$ and 
${\rm{Re}} \chi_\p (q_x,\pi;0)$, are given by 
$2\eta_\theta/\pi v_\theta$ and $2 / \pi v_F$, respectively,   
 while  ${\rm{Re}} \chi_\p (q_x,0;0)$ and 
${\rm{Re}} \chi_{\rho (\p)} (q_x,\pi;0)$ are calculated as follows,  
\begin{eqnarray}
{\rm{Re}} \chi_\p (q_x,0;0) 
&=& \frac{2}{\pi v_\phi} 
\left\{
1 - \frac{2 \Delta_s^2}{v_\phi q_x \sqrt{(v_\phi q_x)^2 + 4 \Delta_s^2}}
\ln \frac{| v_\phi q_x + \sqrt{(v_\phi q_x)^2 + 4 \Delta_s^2}|}
{| v_\phi q_x  - \sqrt{(v_\phi q_x)^2 + 4 \Delta_s^2}|}
\right\} \virg
\label{eqn:rechis0}
\\
{\rm{Re}} \chi_\rho (q_x,\pi;0) 
&=& \frac{2}{\pi v_F}
\sum_{\nu = \pm}
\left\{
\frac{1}{2}
- \frac{ \tiD^2}{\xi_\nu \sqrt{ \xi_\nu^2 + 4 \tiD^2}}
  \ln \frac{| \xi_\nu + \sqrt{ \xi_\nu^2 + 4 \tiD^2}|}
           {| \xi_\nu - \sqrt{ \xi_\nu^2 + 4 \tiD^2}|}
\right\}_{\xi_\nu = v_F(\nu q_x + q_0)}
 \point 
\label{eqn:rechir}
\end{eqnarray}
 In Fig.\ref{fig:2},  
 we show the real parts of the response functions  
  which are normalized by $2/\pi v_F$.  
 For $q_y = 0$, 
 ${\rm Re} \chi_\rho (q_x,0;0)$ as a function of 
 $q_x$ remains   constant.  
 The quantity  ${\rm Re} \chi_\p (q_x,0;0)$ is reduced around $q_x = 0$ owing to the spin gap where  
 the limiting behavior of ${\rm Re} \chi_\p (q_x,0;0)$ for $|q_x| \ll |\Delta_s|/v_\phi$
 is  given by $2/(\pi v_\phi) \times v_\phi^2 q_x^2 / (6 \Delta_s^2)$.
  For $q_y = \pi$, 
  the gaps of the transverse fluctuations, 
$\tiD_s$ and $\tiD_c$, lead to the suppression of 
 ${\rm Re} \chi_\rho (q_x,\pi;0)$ 
 around $q_x = \pm q_0$. 
The  minimum of
 ${\rm Re} \chi_\rho (q_x,\pi;0)$ 
 near $q_x = \pm q_0$   
 comes from 
the separation of Fermi wave vector which is caused by the  
  interchain hopping. 
The  quantity, ${\rm Re} \chi_\rho (q_x,\pi;0)$, 
 for $|q_x \pm q_0| \ll |\tiD|/v_F$ is expressed as 
$2/(\pi v_F) \times \{ 1/2 - (\tiD/4t)^2 \ln (4t/\tiD)^2 + v_F^2 (q_x \pm q_0)^2 / 12 \tiD^2 \}$. 
The quantity  ${\rm Re} \chi_\p (q_x,\pi;0)$ as a function of $q_x$ 
 becomes flat  
  due to the fact that the effect of $\tiD_c$ and $\tiD_s$ 
compensate each other in Eq.(\ref{eqn:chipi}).  

\begin{figure}
\centerline{\epsfysize=11.0cm\epsfbox{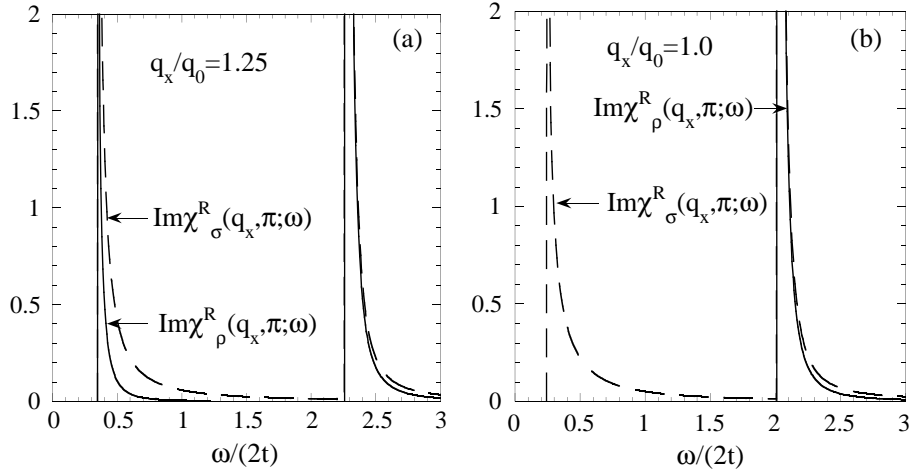}}
\vspace{-4.2cm}
\caption{
 The $\omega/(2t)$-dependence of
 ${\rm Im} \chi^R_\rho(q_x,\pi;\omega)$ and 
 ${\rm Im} \chi^R_\p(q_x,\pi;\omega)$
for   $q_x/q_0 = 1.25$ (a) 
  and $q_x/q_0 = 1.0$ (b),
  which are normalized by $2 / \pi v_F$.   
 The solid (dashed) curve represents ${\rm Im} \chi^R_\rho(q_x,\pi;\omega)$
 (${\rm Im} \chi^R_\p(q_x,\pi;\omega)$). 
 The parameters are the same as those in Fig.2.
 }
\label{fig:3}
\end{figure}

 Now we examine the $\omega$-dependence of the imaginary 
 parts of Eq.(\ref{eqn:chipi}) 
 which represents the spectral weight 
 for charge and spin fluctuations with $q_y = \pi$.  
In Fig.\ref{fig:3},  
 the   quantities   ${\rm Im} \chi^R_\rho (q_x,\pi;\omega)$ and ${\rm Im} \chi^R_\p (q_x,\pi;\omega)$, which are 
normalized by $2/\pi v_F$, 
 are  shown with the fixed $q_x / q_0 =$ 1.25 ( Fig.\ref{fig:3}(a) )
and 1.0 ( Fig.\ref{fig:3}(b) ).  
 These quantities diverge  at two  locations 
 given by  
$\omega_{L(H)}/(2t) = \sqrt{\{1 -(+) q_x/q_0\}^2 + 4 \{\tiD/(2t)\}^2}$,  which are ascribed  to the separation of the Fermi wave vector  
by the interchain hopping. 
 These singularities are also found 
  in the absence of the interaction, 
where the corresponding response function 
 becomes 
$ \chi^R_{\rho(\sigma)} (q_x,\pi;\omega) \rightarrow 
 \chi^R (q_x,\pi;\omega)$ with 
\begin{eqnarray}
\chi^R (q_x,\pi;\omega)
 = \frac{1}{2\pi} \sum_{\nu = \pm}
\Bigg\{
  \frac{\nu q_x + q_0}{v_F(\nu q_x + q_0) - \omega - \im \de} 
+ \frac{\nu q_x + q_0}{v_F(\nu q_x + q_0) + \omega + \im \de} 
\Bigg\} \point
\label{eqn:chifree}
\end{eqnarray}
 The weight of Eq.(\ref{eqn:chipi}) 
 appears in the small  region of the energy 
  higher  than $\omega_{L(H)}$,   
 while Eq.(\ref{eqn:chifree}) shows 
  the weight with  a delta function 
 in the absence of the interaction. 
 The  location of $\omega$  corresponding to  
the singularity depends on  the magnitude of  the gap.
It is worthy to note that the lower peak  of 
 the response function  shows the dominant behavior 
 of the spin fluctuation 
 compared with the charge fluctuation. 
 For $q_x = q_0$,
 the weight  at the lower energy 
 is given only by the spin fluctuation, i.e., 
  the charge fluctuation is absent 
  as is shown in Fig.\ref{fig:3}(b).
\begin{figure}
\centerline{\epsfysize=11.0cm\epsfbox{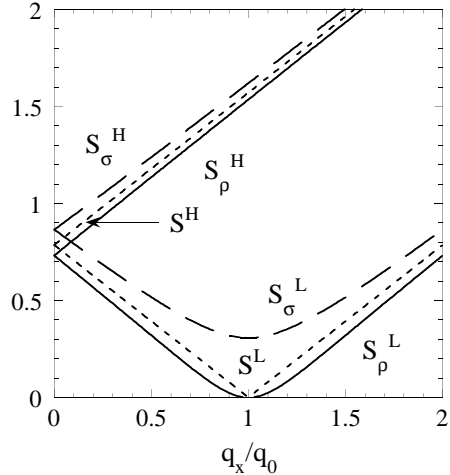}}
\vspace{-4.2cm}
\caption{
 The total weights,  
$S_{\rho(\p)}^L$ and 
$S_{\rho(\p)}^H$ 
as a function of $q_x/q_0$.  
The dotted lines, $S^L$ and $S^H$ are those in the absence of the interaction.
 The parameters are the same as those in Fig.2.
 }
\label{fig:4}
\end{figure}

 In order to examine such a notable fact, we evaluate the weights 
 by dividing  Eq.(\ref{eqn:chipi})  
 into two parts, i.e.,   
${\rm Im} \chi^{R}_{\rho(\p)}(q_x,\pi;\omega) = 
{\rm Im} \chi^{R,L}_{\rho(\p)}(q_x,\pi;\omega) +
{\rm Im} \chi^{R,H}_{\rho(\p)}(q_x,\pi;\omega)$, 
 where 
 ${\rm Im} \chi^{R,L}_{\rho(\p)}(q_x,\pi;\omega)$  
( ${\rm Im} \chi^{R,H}_{\rho(\p)}(q_x,\pi;\omega)$ )
 comes from the part with $\nu = - (+)$, and leads to 
the singularity near     
 $\omega = \omega_L$ ($\omega_H$). 
 In Fig.\ref{fig:4}, 
we show the total weight of 
${\rm Im} \chi^{R,L}_{\rho(\p)}(q_x,\pi;\omega)$ and 
 ${\rm Im} \chi^{R,H}_{\rho(\p)}(q_x,\pi;\omega)$ 
 which are defined by  
\begin{eqnarray}
S_{\rho(\p)}^L &=& \frac{\pi v_F}{4t} \int_0^\infty \d \omega
{\rm Im} \chi^{R,L}_{\rho(\p)}(q_x,\pi;\omega) \virg \\
\label{sumL}                       
S_{\rho(\p)}^H &=& \frac{\pi v_F}{4t} \int_0^\infty \d \omega
{\rm Im} \chi^{R,H}_{\rho(\p)}(q_x,\pi;\omega) \point
\label{sumH}
\end{eqnarray}
 Solid and dashed curves denote 
 $S_{\rho}^{L,H}$ and 
 $S_{\sigma}^{L,H}$, respectively while   
$S^L$ and $S^H$ shown by the dotted lines are the total weights 
  in the absence of the interaction. 
 The deviation of 
 $S_{\rho}^{L,H}$ from $S^{L,H}$ is opposite to that of 
 $S_{\sigma}^{L,H}$.  
 The fact that $S_{\sigma}^L \gg S_{\rho}^L$ around $q_x \sim q_0$
 shows the clear evidence that
 the spin fluctuation  with low energy 
dominates  the charge fluctuation.  
The results indicate that the low-lying excitation with $q_y = \pi$
 in case of  the repulsive intrachain interaction   
 is  determined mainly by the spin degree of freedom. 

\section{Summary and Discussion}
In the present paper, 
 we investigated the  response functions for
  charge and spin densities 
 in  the system of the two chains of the electron system 
with the repulsive intrachain interaction  
 by using the method of the effective Hamiltonian based on the 
 renormalization group analysis. 

In the several non-linear terms of the bosonized Hamiltonian, 
we replaced the terms with the finite expectation value by the 
averaged one and utilized  the mean-field approximation 
 in order to estimate the gap. 
Such an approximation seems  effective in the sense that 
the method gives the results  consistent  
with the ones derived from the 
renormalization group analysis.  
In addition, we used the approximation 
 $\eta_\phi \rightarrow 1$ to  express 
the mean-field Hamiltonian by the new Fermion field. 
Such a setting does not change qualitatively the behavior 
of the total spin fluctuation, \ie the spin gap appears. 
We note that such a treatment of $\eta_\phi = 1$ breaks 
$SU(2)$ symmetry which is preserved in Eq.(1).  
In fact, the formulas for the spin susceptibilities, 
$\chi_{xx}^R(q_x,0;\omega)$ 
and $\chi_{xx}^R(q_x,\pi;\omega)$, which are  
the response functions for spin with the $x$-direction, 
are different from Eqs.(16) and (17) 
in the sense that  
$\chi_{xx}^R(q_x,0;\omega)$ 
($\chi_{xx}^R(q_x,\pi;\omega)$) contains the gaps of both
$\Delta_s$ and $\tiD_s$ ($\Delta_s$ and $\tiD_c$). 
However, the behaviors of 
$\chi_{xx}^R(q_x,0;\omega)$ and 
$\chi_{xx}^R(q_x,\pi;\omega)$  
are approximately the same  as those of Eqs.(16) and (17), respectively.  
For the real part of the susceptibility,      
${\rm Re} \chi_{xx}^R(q_x,0;0)$ 
is strongly suppressed near $q_x = 0$
(${\rm Re} \chi_{xx}^R(q_x,0;0) \sim 0.05$ in unit of $2/\pi v_F$ 
for the parameters in Fig.2)
 and
${\rm Re} \chi_{xx}^R(q_x,\pi;0)$ does not show
the remarkable suppression near $q_x = q_0$.
In addition, 
there are the two peaks in ${\rm Im} \chi_{xx}^R(q_x,\pi;\omega)$ 
as a function of $\omega$ even at $q_x = q_0$
and the total weight of the lower peak is larger 
than that of the non-interacting case. 
Thus the present treatment
leads to results qualitatively reasonable for spin fluctuations  
though $SU(2)$ symmetry is broken.
\begin{figure}
\centerline{\epsfysize=11.0cm\epsfbox{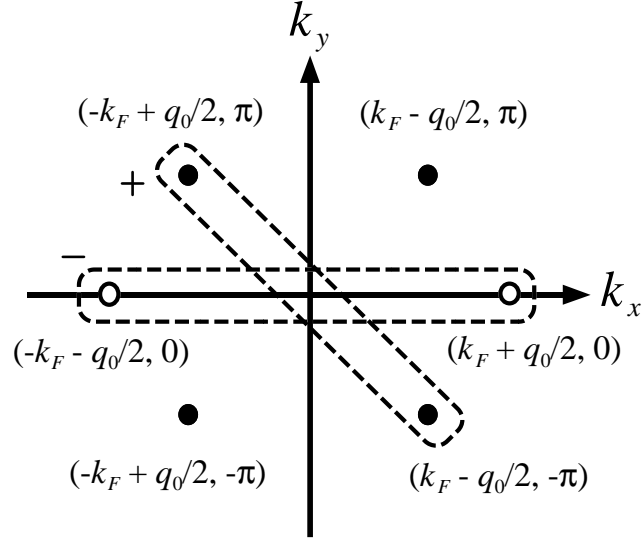}}
\vspace{-3.2cm}
\caption{
The pairing in the momentum space $(k_x,k_y)$ is shown schematically by the enclosed-dotted curve. 
The black ( white ) circles show the Fermi points of the upper ( lower ) band.   
The Fermi points with $k_y = -\pi$ are equivalent to 
those with $k_y = \pi$. 
 The sign  $+$ and $-$ expresses the fact that   
 the order parameters of the superconducting state is out of phase.  
 }
\label{fig:5}
\end{figure}

From the imaginary part of the response functions with $q_y = \pi$, 
we concluded that 
 among the fluctuations with $q_y = \pi$, 
the spin degree of freedom is dominant 
compared to the charge one. 
The  excitation of  spin degree of freedom  with $q_y = \pi$ and 
 low energy seems to be closely related to superconductivity of the two chains. 
In the present superconducting state, 
the pair of the electrons is formed between two chains with 
 {\it in phase}. 
This 
 corresponds to the intraband pairing with 
 {\it out of phase} in the momentum space $(k_x,k_y)$
\cite{Fabrizio-Parola-Tosatti,Fabrizio,Schulz,Yoshioka-Suzumura-4,Yoshioka-Suzumura-5} 
 where the Fermi points are given by 
 $(\pm(k_F+q_0/2),0)$ for the lower band and 
 $(\pm(k_F-q_0/2),0)$ for the upper band, respectively. 
 The  superconducting state 
  is explained  in 
  Fig.\ref{fig:5} where   the pairing of the electrons 
  around the Fermi points  are expressed by the enclosed-dashed curves.
  By noting that 
$\chi_{\rho(\p)} (q_x,\pi;\im \on) = (-1/\pi) \int_{-\infty}^{\infty} 
\d \omega
 {\rm Im} \chi^R_{\rho(\p)} 
(q_x,\pi;\omega) /( \im \on - \omega)$, 
it is considered that  
the fluctuations connecting these two kinds of pairs,  
$\chi_{\rho(\p)} (q_x,\pi;\im \on)$ with $q_x \simeq \pm q_0$
  act as  the attractive force \cite{Monthoux-Pines}.
Since the spin degree of freedom in such fluctuation is dominant 
compared to the charge fluctuation, 
it is possible that the spin fluctuation with $q_y = \pi$
 results in the superconductivity of the two chains. 
 Also in the organic conductors, ${\rm (TMTSF)_2 X}$,  
 it is argued  that the pair is formed between  the chains 
from the theoretical analysis  
\cite{Hasegawa-Fukuyama} of the NMR relaxation rate of ${\rm (TMTSF)_2 
ClO_4}$  
\cite{Takigawa-Yasuoka-Saito}. 
Therefore the origin of the superconductivity of the organic conductor 
seems to be  the spin fluctuation with $q_y = \pi$.

\section*{Acknowledgment} 
 This work was partially supported 
 by  the Grant-in-Aid for Scientific  Research 
 on the priority area, Novel Electronic States in Molecular Conductors,
 from the Ministry of Education, Science and Culture, Japan.



\begin{thebibliography}{99}
%
\bibitem{Dagotto-Rice-Rev}
For review, E. Dagotto and T. M. Rice, 
\jo{Science}{271}{1996}{618}.
%
\bibitem{Dagotto-Riera-Scalapino}
E. Dagotto, J. Riera and D. J. Scalapino,
\jo{\PRB}{45}{1992}{5744}.
%
\bibitem{Fabrizio-Parola-Tosatti}
M. Fabrizio, A. Parola and E. Tosatti, 	 
\jo{\PRB}{46}{1992}{3159}.
%
\bibitem{Fabrizio}
M. Fabrizio, 	 
\jo{\PRB}{48}{1993}{15838}.
%
%
\bibitem{Shimoi-Yamaji-Yanagisawa}
K. Yamaji and Y. Shimoi, 	 
\jo{\PHC}{222}{1994}{349}; 
T. Yanagisawa,	Y. Shimoi and K. Yamaji, 
\jo{\PRB}{52}{1995}{R3861}. 
%
\bibitem{Nagaosa}
N. Nagaosa, 
\jo{\SSC}{94}{1995}{495}; 
N. Nagaosa and M. Oshikawa
\jo{\JPSJ}{65}{1996}{2241}. 
%
\bibitem{Schulz}
H. J. Schulz, 
\jo{\PRB}{53}{1996}{R2959}, 
1996 cond-mat preprint 9605075. 
%
\bibitem{Balents-Fisher}
L. Balents and M. P. A. Fisher, 
\jo{\PRB}{53}{1996}{12133}.
%
\bibitem{Shelton-Tsvelik}
D. G. Shelton and A. M. Tsvelik:	 
\jo{\PRB}{53}{1996}{14036}.  
%
%
%
\bibitem{Johnston-Johnston}
D. C. Johnston, J. W. Johnston, D. P. Goshorn and A. P. Jacobson, 
\jo{\PRB}{35}{1987}{219}.
%
\bibitem{Hiroi}
Z. Hiroi, M. Azuma, M. Takano and Y. Bando,
\jo{\JSSC}{95}{1991}{230}.
%
%
\bibitem{Senachal}
D. S\'en\'echal
\jo{\PRB}{52}{1995}{15319}.
%
\bibitem{Uehara-Nagata-Akimitsu-Takahashi-Mori-Kinoshita}
M. Uehara, T. Nagata, J. Akimitsu, H. Takahashi, N. M$\hat{\rm o}$ri 
and K. Kinoshita, 
\jo{\JPSJ}{65}{1996}{2764}.
%
\bibitem{Jerome}
 D. J\'erome, 
\jo{\MCLC}{79}{1982}{155}.
%
\bibitem{TMTSF-rev}
 For reviews, 
D. J\'erome and H. J. Schulz, 
 \jo{\ADV}{31}{1982}{299}; 
 T. Ishiguro and K. Yamaji, 	 
 {\it Organic Superconductors}, 
  Springer Series in Solid-State Sciences  
 (Springer-Verlag, Berlin, 1990), Vol.88.
%
\bibitem{Yoshioka-Suzumura-4}
H. Yoshioka and Y. Suzumura, 
\jo{\PRB}{54}{1996}{9328}.
%
\bibitem{relevant-condition}
The condition is given in ref.\cite{Nagaosa}
and $\max (t_{\parallel},t) \de  \gg J_\bot$ where 
$t_{\parallel}$ ($t$), $\de$ and $J_\bot$ are the intrachain (interchain) hopping energy,  
the hole-concentration 
 and 
the interchain exchange energy, respectively.   
%
\bibitem{Yoshioka-Suzumura-5}
H. Yoshioka and Y. Suzumura,
\jo{\JLTP}{109}{1997}{49}.   
%
\bibitem{Solyom}
 J. S\'olyom,
\jo{\ADV}{28}{1979}{201}.	 
%
\bibitem{notation}
Correspondences between the present notation of the phase variables
and that by Schulz\cite{Schulz} are as follows, 
$\theta_+ = \sqrt{2} \phi_{\rho +}$,
$\theta_- = - \sqrt{2} \theta_{\rho +}$,
$\phi_+ = \sqrt{2} \phi_{\p +}$,
$\phi_- = - \sqrt{2} \theta_{\p +}$,
$\titheta_+ = \sqrt{2} \phi_{\rho -}$,
$\titheta_- = - \sqrt{2} \theta_{\rho -}$,
$\tiphi_+ = \sqrt{2} \phi_{\p -}$ and 
$\tiphi_- = - \sqrt{2} \theta_{\p -}$.
%
\bibitem{Finkelstein-Larkin}
A. M. Finkelstein and A. I. Larkin, 
\jo{\PRB}{47}{1993}{10461}.
%
\bibitem{Luther-Emery}
A. Luther and V. J. Emery 
\jo{\PRL}{57}{1974}{589}. 
%
%
%
\bibitem{Monthoux-Pines}
P. Monthoux and D. Pines
\jo{\PRB}{47}{1993}{6069}.
%
\bibitem{Hasegawa-Fukuyama}
Y. Hasegawa and H. Fukuyama,  
\jo{\JPSJ}{56}{1987}{877}.
%
\bibitem{Takigawa-Yasuoka-Saito}
M. Takigawa, H. Yasuoka and G. Saito,  
\jo{\JPSJ}{56}{1987}{873}.
%
%

\end{thebibliography}
\end{document}